\documentclass[aps,prd,superscriptaddress,showpacs,amsmath,amssymb]{revtex4}
\usepackage{graphicx,epsf}
\usepackage{amsfonts}
\usepackage{amssymb}
\usepackage[dvips]{color}
\definecolor{red}{rgb}{1,0,0}
\definecolor{green}{rgb}{0,1,0}

\newcommand{\be}{\begin{eqnarray}}
\newcommand{\ee}{\end{eqnarray}}

\begin{document}

\title{Adiabatic \& non-adiabatic perturbation theory for coherence vector description of neutrino oscillations}

\author{Sebastian Hollenberg}
\email{sebastian.hollenberg@tu-dortmund.de}
\author{Heinrich P\"as}
\email{heinrich.paes@tu-dortmund.de}
\affiliation{Fakult\"at f\"ur Physik, Technische Universit\"at Dortmund, D-44221 Dortmund, Germany}

\begin{abstract}
The standard wave function approach for the treatment of neutrino oscillations fails in situations where quantum
ensembles at a finite temperature with or without an interacting background plasma are encountered. 
As a first step to treat such phenomena in a novel way, we propose a unified approach to both 
adiabatic and non-adiabatic two-flavor oscillations in neutrino ensembles with finite temperature and generic
(e.g. matter) potentials. Neglecting effects of ensemble decoherence for now we study the evolution of a neutrino ensemble governed by the associated Quantum Kinetic Equations, which apply to systems with finite temperature. The Quantum Kinetic Equations are solved formally using the Magnus expansion and it is shown that a convenient choice of the quantum mechanical picture (e.g. the interaction picture) reveals suitable parameters to characterize the physics of the underlying system (e.g. an effective oscillation length).
It is understood that this method also provides a promising starting point for the treatment of the more general
case in which decoherence is taken into account.
\end{abstract}
\preprint{DO-TH-??}

\maketitle

\section{Introduction}\label{intro}
Ever since the conjecture that neutrinos have mass and thus might be subject to flavor oscillations
\cite{Gribov:1968kq} there has been a thriving interest in this very phenomenon which clearly
indicates and gives rise to speculations as to how to describe physics beyond the electro-weak standard model.
\par
A convenient and established way to deal with neutrino flavor oscillations is to encode this effect in a Hamiltonian formulation in which the oscillatory behavior is captured in a Schr\"odinger-like equation for a wave function in neutrino flavor space. This formalism, in principle, applies to an arbitrary number of neutrino generations and is also capable of incorporating medium effects on neutrino propagation such as coherent elastic forward scattering in, e.g., stellar matter \cite{Wolfenstein:1977ue}. It was soon realized that the Hamiltonian formalism for
neutrino oscillations can be given a geometrical interpretation in $N^2-1$-dimensional flavor space for $N$ neutrino flavors \cite{Kim:1987bv}. This approach to neutrino oscillations sees equations of motion for a
coherence vector~\footnote{Different names exist for the coherence vector depending on the area of physics it is used in: Bloch vector is most common in solid state physics, whereas polarization or coherence vector is more common in
particle physics applications.} in that the Schr\"odinger-like equation of motion can be rephrased as a gyroscope
equation, i.e. a formal equivalent to e.g. the precession of a magnetic moment in an external magnetic field. Besides its apparant usefulness when it comes to picture neutrino oscillations there is also a purely formal merit to the
gyroscope-type equations in that they are introduced by means of decomposing the Hamiltonian in terms of
the generators of the associated $SU(N)$, e.g. the Pauli matrices for a two-flavor system. 
\par
This decomposition procedure is also most convenient when the notion of a wave function is not suitable any more to describe the
physics of neutrino oscillations. A typical situation in which the breakdown of the wave function formalism is expected
are quantum ensembles with a finite temperature or neutrino ensembles with a finite temperature and an interacting
background plasma. The latter situation is encountered in the early universe prior to Big Bang Nucleosythesis. The crucial point in such an environment is the breaking of coherence due to the small mean free path of neutrinos at high
temperatures. The other important modification results from the fact that neutrino oscillations in the early universe can alter the lepton asymmetry, which in turn contributes to the refraction index of the primeval plasma
rendering the equations of motion nonlinear. It is essentially for those two reasons that the wave function formalism must fail in describing neutrino oscillation phenomena now. The appropriate description is then given by 
the density matrix formalism. The density matrix of the neutrino ensemble obeys a von Neumann equation and the different contributions to the effective Hamiltonian are given by collisions (non-forward scattering), with particles from the background medium, which introduce decoherence. On the other hand coherent oscillations are governed by a matter-dependent effective Hamiltonian which comes about via coherent forward scattering processes of neutrinos off
the background particles \cite{McKellar:1992ja, Sigl:1992fn}.
\par
The dynamics of the neutrino ensemble are determined by the Quantum Kinetic Equations (QKE) which present a generalization of the Pauli-Boltzmann equations. The former evolve quantum amplitudes as is indispensable
if a consistent description of particle oscillation phenomena, which are inherently nonclassical, is sought. The
Pauli-Boltzmann approach on the other hand evolves probabilities rather than amplitudes; this procedure is essentially
classical since quantum mechanics only enters the problem when it comes to calculating cross sections for the various
possible reaction channels. The resultant Quantum Rate Equations (QRE) are inappropriate when neutrino oscillations
occur. Thus, in order to obtain the QKE the full density matrix for all particles in the plasma is evolved forward
in time by means of the $S$ matrix and tracing over all degrees of freedom other than the neutrinos under
consideration yields the equation of motion for the system's density matrix, the QKE, which do reduce to QRE in the
appropriate limit \cite{McKellar:1992ja}. The variable of interest in the QKE is the one-body reduced momentum-dependent density operator, which is conveniently decomposed in terms of the generators of the associated 
$SU(N)$.
\par
In the forthcomig analysis we account for a brief motivation on how to obtain the QKE from the density matrix
formalism in the case of coherent forward-scattering (which also dominates the bulk of the studies to follow) and
how to relate the solutions to the QKE, i.e. the coherence vector, to physical observables in section \ref{notation}. Moreover, we introduce the Magnus expansion, which allows for an analytic, yet approximative, solution to the underlying QKE. A short discussion of the convergence properties of the Magnus expansion proves useful to single out
suitable quantities to describe the physics at hand. In section \ref{ad} we unfold a systematic way to develop
an adiabatic perturbation theory starting from the QKE for a two-flavor neutrino ensemble with generic potentials.
We show that the Magnus expansion not only allows to analytically solve the QKE in a perturbative way, but 
can also serve to motivate the definition of physical quantities, such as an effective mixing angle or an adiabaticity parameter. In order to isolate a convenient perturbation parameter we perform different changes of the quantum mechanical picture (e.g. into the interaction picture) for the QKE. Hence, a suitable succession of bases changes can improve the convergence properties of the expansion. We solve the QKE for the coherence vector to first order in the Magnus expansion. The perturbation parameter in the adiabatic perturbation theory can then be used to identify a suitable expansion parameter for the non-adiabatic case. Once this parameter is identified the non-adiabatic perturbation theory can be treated on the same ground as the adiabatic one, i.e. singling out an appropriate expansion parameter by changing the quantum mechanical picture and solving the QKE in the resulting representation by means of the Magnus 
expansion. This non-adiabatic perturbation theory is developed in section \ref{nad}. In section \ref{limit} we comment
on the integrals appearing in both the adiabatic and non-adiabatic perturbation theories and show how the latter lead to the correct limits such as vacuum neutrino oscillations and the slab model approximation to non-adiabatic transitions in the MSW framework. We sketch how to extend the developed perturbation theory to higher orders in the 
Magnus expansion. Moreover, we elaborate on possible extensions of the theory introduced to
scenarios including decoherence in the ensemble due to collision processes with particles from a background plasma.
It turns out that the continuation of the perturbation theory is somewhat nontrivial and deserves a more careful study.

\section{Notation and mathematical tool box}\label{notation}

The purpose of our analysis is to describe the evolution of a two-flavor neutrino ensemble 
with generic potentials at a finite temperature. To this end we commence our considerations by a close inspection
of the underlying QKE, which in this context can be readily derived from the density matrix formalism for neutrino
oscillations. We remark that the latter has to be used to correctly deal with quantum systems at finite temperature 
as well as in situations where loss of coherence becomes important. The wave function formalism ceases to provide
an appropriate handle for such systems. However, it is not until section \ref{limitbeyond} that we discuss an
extension of the formalism to be discussed shortly to systems with a background plasma, hence decohering collisions, present. 
\par
The density matrix $\rho(p,t)$ obeys a von Neumann equation
\be
   \dot{\rho}(p,t) = -i \left[H(p,t),\rho(p,t)\right],
\ee
where $p$ is the neutrino four-momentum, $H$ a generic Hamiltonian for the system. 
Since we are dealing with a two-flavor
system we can now decompose both the density matrix and the Hamiltonian in terms of the generators of the
associated $SU(2)$, namely the Pauli matrices $\sigma^i$. This yields
\be
   \rho &=& \frac{1}{2}\text{Tr}\rho \left[ 1 + \text{Tr}\left(\rho \sigma^i\right) \sigma^i\right]
        \equiv \frac{1}{2} P_0 \left[ 1 + \vec{P} \vec{\sigma}\right], \\
   H &=& \frac{1}{2}\text{Tr}H \left[ 1 + \text{Tr}\left(H \sigma^i\right) \sigma^i\right]
        \equiv \frac{1}{2} V_0 \left[ 1 + \vec{V} \vec{\sigma}\right],
\ee
where repeated indices are to be summed over and the obvious identifications $P_0 = \text{Tr}\rho$, 
$V_0 = \text{Tr}H$, $\vec{P} = \text{Tr}\left(\rho\vec{\sigma}\right)$, 
$\vec{V} = \text{Tr}\left(H\vec{\sigma}\right)$ and $\vec{\sigma} = \left(\sigma^x, \sigma^y, \sigma^z\right)$ 
have been made. Here $\vec{P}$ is the so-called coherence vector \cite{Kimura}. 
Making use of this notation it is straightforward to recast the von Neumann equations according to
\be
   \dot{P}_i = \left[- V_0 \varepsilon_{ilk} V^k\right] P^l.
\ee
Hence we can identify
\be
   S_{il} \equiv - V_0 \varepsilon_{ilk} V^k \qquad \text{or} \qquad 
   S = V_0 \begin{pmatrix}
            0 & -V_z & V_y \\ V_z & 0 & -V_x \\ -V_y & V_x & 0
           \end{pmatrix}
\ee
as the evolution matrix of the neutrino ensemble~\footnote{Note that the
matrix $S$ as defined here does not coincide with the $S$ matrix of the plasma mentioned before. Since we will not
refer to the latter explicitly any more no confusion should arise from this notation.}. Put another way, the QKE can now be 
written~\footnote{Note, that this equation can also be given as 
$\dot{\vec{P}} = V_0 ~ \vec{V} \times \vec{P}$ using the cross product in three dimensional Euclidean space. We
remark that this gyroscope-type equation for neutrino oscillations presents a convenient way to interpret neutrino oscillations geometrically. However, when it comes to solving the QKE it is appropriate to treat them as a 
non-autonomous system of coupled differential equations as will be seen in due course.}
as
\be
   \frac{\mathrm d}{\mathrm d t}\vec{P}(t) = S(t)\vec{P}(t) \label{QKE}.
\ee
The entries of the effective potential vector $\vec{V} = (V_x, V_y, V_z)$ can also be readily obtained via
\be
   V_0 V_x &\equiv& \beta = 2 \text{Re} H_{12} = \frac{\Delta m^2}{2p} \sin2\theta_0, \\
   V_0 V_y &=& 2 \text{Im} H_{12} = 0, \\
   V_0 V_z &\equiv& \lambda = \left(H_{11} - H_{22}\right) = -\frac{\Delta m^2}{2p} \cos2\theta_0 
                                                                       + V_{\alpha},
\ee
where $H_{ij}$ are the elements of the Hamiltonian $H$, $\Delta m^2$ is the mass-squared difference of the two
neutrino states, $\theta_0$ is the associated vacuum mixing angle between the two flavors, which we shall denote as
$\nu_{a}$ and $\nu_{b}$ for definiteness. $V_{\alpha}$ is the difference of potential terms affecting $\nu_a$ and
$\nu_b$ respectively. The last equality has been obtained using the $2 \times 2$ neutrino oscillation Hamiltonian
in flavor space.
\par
In the coherence vector description the expectation values of the generators of the associated $SU(2)$ are 
promoted to observables of interest. All information about the system can thus, in principle, be extracted from
a solution to Eq.(\ref{QKE}) for $\vec{P}$. Some comments related to this issue are in order:
In the one-particle interpretation the diagonal entries of the density matrix simply give the probability to find 
the system in one or the other state, i.e.
\be
   \text{prob}\left(\nu_a \to \nu_a\right) &=& \frac{1}{2}P_0 \left[1 + P_z\right], \label{probaa} \\
   \text{prob}\left(\nu_a \to \nu_b\right) &=& \frac{1}{2}P_0 \left[1 - P_z\right] \label{probab}.
\ee
In the ensemble interpretation of the density matrix the diagonal entries give the relative number densities for 
the different neutrino flavors normalized to the equilibrium Fermi-Dirac number distribution at zero chemical potential $\mu$ according to
\be
   N_a(p) &=& \frac{1}{2} P_0 \left[1 + P_z\right] N^{\text{EQ}}(p,0), \\
   N_b(p) &=& \frac{1}{2} P_0 \left[1 - P_z\right] N^{\text{EQ}}(p,0), \\
   N^{\text{EQ}}(p,\mu) &=& \frac{1}{2\pi^2} \frac{p^2}{1 + e^{\frac{p-\mu}{T}}},
\ee
where $T$ is the temperature of the ensemble. Those relations are also easily inverted to yield a physical meaning of
both
\be
   P_0 &=& \frac{N_a + N_b}{N^{\text{EQ}}}, \\
   P_z &=& \frac{N_a - N_b}{N_a + N_b}. 
\ee
Hence $P_0$ is connected to conservation of probability and, in a broader context, also lepton number. It is important to note that oscillations merely swap neutrinos from one flavor to another so that $P_0$ does not evolve in time, unless repopulation effects from some background plasma have to be taken into account as is the case, e.g., in the early universe. On the other hand $P_z$ parametrizes the asymmetry of the system that is the excess of $\nu_a$
over $\nu_b$. The latter fact also motivates the way of speaking in which $P_x$, $P_y$ are called coherences encoding
the amount of decoherence in the system. Therefore the evolution of $P_z$ is of special interest in most applications.
\par\noindent
\newline
Our next concern is to provide mathematical means to approximately solve the 
differential equations~(\ref{QKE}). In order to do so we rewrite the QKE supplied with
an initial condition as
\be
   \frac{\partial}{\partial t} \vec{P}(t) = S(t)\vec{P}(t), \qquad \vec{P}^i = \vec{P}(t_0) \label{QKE2}
\ee
and notice that we are dealing with a non-autonomous set of linear differential equations.
If the matrix $S$ did not 
depend on time, the corresponding differential equation could be readily solved by taking the
matrix exponential $\text{exp}[S(t-t_0)]$ and writing $\vec{P}(t) = \text{exp}[S(t-t_0)] \vec{P}^i$. 
It is then fair to ask whether the solution to Eq.(\ref{QKE2})
can always be written as an exponential via
\be
   \vec{P}(t) = \text{exp}[\Omega(t,t_0)]~\vec{P}^i.
\ee
A method for finding such a true exponential solution~\footnote{Note that this is in contrast to the 
Dyson series solution which is frequently encountered in Quantum Field Theory applications. The latter
relies on time-ordered products of operators which sees some disadvantages if the series has to be truncated, e.g.
that unitarity is only restored up to the order considered, i.e. by explicitly neglecting higher order contributions.}
has indeed been established under the label Magnus expansion \cite{Magnus:1954, Blanes:2009}. The Magnus operator $\Omega = \ln S$ satisfies its 
own differential equation which in turn is solved by a series expansion $\Omega = \sum_{n=1}^{\infty} \Omega_n(t)$,
where each Magnus approximant $\Omega_n(t)$ is small in an appropriate sense. The smallness of each Magnus
approximant then determines the convergence properties of the expansion. A general theorem states that for a
differential equation (\ref{QKE2}) defined in a Hilbert space with a bounded operator $S$ the series converges
in the interval $t \in [t_0,t_c[$ such that 
\be
   \int_{t_0}^{t_c}\mathrm d\tau  ~\vert\vert S(\tau)\vert\vert < \pi \label{convcond},
\ee
where $\vert\vert ~.~ \vert\vert$ is a matrix norm. As can be inferred from this condition the convergence
properties of the Magnus expansion can be improved by means of a change of the quantum mechanical picture and 
thus improving the convergence properties of the expansion can be rephrased as minimizing the matrix norm of 
the evolution matrix.
\par
The first two Magnus approximants assume a form
\be
   \Omega_1(t) &=& \int_{t_0}^{t}\mathrm d \tau ~ S(\tau), \\
   \Omega_2(t) &=& \frac{1}{2}\int_{t_0}^{t} \mathrm d t_1 \int_{t_0}^{t_1} \mathrm d t_2 ~ [S(t_1),S(t_2)]
\ee
and various methods have been worked out to calculate higher order terms 
\cite{Klarsfeld:1989zz}. All such terms contain nested 
commutators of $S$ evaluated at different times from which it is readily seen that the Magnus expansion gives
the exact result (the matrix exponential for a time-independent matrix $S$) already in first order, if the
matrix $S$ commutes with the matrix obtained by integrating $S$ over a certain time interval. Note that each
approximant $\Omega_n$ adopts the same properties as the matrix $S$. If $S$ was (anti-)Hermitian, so would
be the exponentials of all approximants; this means that unitarity of the solution will be preserved in each order of the perturbation expansion seperately.
\par
The convenience of the Magnus expansion is now apparent: In lieu of a calculation of eigenvalues and eigenvectors 
of a matrix it poses the problem of calculating a matrix exponential. The latter problem is
usually solvable for $3 \times 3$ matrices. Note, however, that the dimension of the evolution matrix
corresponds to the number of generators of the associated $SU(N)$, i.e. for an $N$-dimensional system
the dimension of the evolution matrix is $N^2-1$. So for the case of three interacting neutrinos 
one already has to choose between
diagonalizing a $8 \times 8$ matrix or calculating its exponential. Both problems are intricate.
\par
Another nice feature of the Magnus expansion is that it provides a clear description of how to improve
the approximation by going to higher order terms. It also reproduces exact solutions to the QKE in the appropriate limits as will be seen later, but is also apparent at this stage of our analysis by inspecting the form
of the Magnus approximants. Moreover, the condition for its convergence appears
as an integral condition on the norm of the evolution matrix. This condition can be used as a cross
check for whether changing from one quantum mechanical picture to another improves convergence properties
of the expansion or not. We will invoke this criterion in the upcoming analysis and as it turns out, it
can even supply some physical insight.

\section{Adiabatic perturbation theory}\label{ad}

Having established QKE of the form
\be
   \frac{\partial}{\partial t}\vec{P}(t) = S(t)\vec{P}(t), \qquad S(t) 
                                         = \begin{pmatrix}
                                            0 & -\lambda(t) & 0 \\ 
                                            \lambda(t) & 0 & -\beta(t) \\
                                            0 & \beta(t) & 0
                                           \end{pmatrix} \label{QKEII}
\ee
it is easy to calculate the matrix norm~\footnote{We are dealing with a finite dimensional vector
space and hence all matrix norms are equivalent. We choose to work with the unitarily invariant Frobenius
norm $\vert\vert S \vert\vert_{\mathrm F}^2 = \text{Tr} \left(S^{\dagger}S\right)$.} according to
\be
   \vert\vert S \vert\vert_{\mathrm F}^2 = \text{Tr} \left(S^{\dagger}S\right) = 2 \omega^2_{\text{eff}}.
\ee
It is straightforward to show that the effective oscillation length of the system is indeed given by
\be
   \frac{2\pi}{l^{\text{eff}}_{\text{osc}}} \equiv \omega_{\text{eff}} = \sqrt{\lambda^2 + \beta^2}.
\ee
At this point of our analysis it might seem academic to calculate the matrix norm of the evolution equation under consideration, but it will be seen shortly that a comparison between matrix norms in different quantum mechanical
pictures can provide physical insight into the nature of the neutrino ensemble at hand. Note, moreover, that we
treat both $\beta$ and $\lambda$ as time-dependent quantities. This generic assumption provides greater freedom
when it comes to adapting the QKE to early universe applications. To this end we notice that $\beta$ scales as
$p^{-1}$ in momentum and, in an expanding universe, this momentum is redshifted and thus depends on time. We shall 
sketch how to treat such situations later on.
\par
Furthermore, in order to get a grasp on how this oscillation length can be understood physically we transform
the QKE to a basis which resembles the commonly encountered mass eigenbasis in the MSW framework. To do so we notice
that there are only non-vanishing contributions to the effective potential vector's $x$ and $z$ component, namely
$V_x$ and $V_z$. Thus it is only sensible to consider a generic time-dependent rotation in the $xz$-plane by an angle
$\Theta(t)$ as
\be
   \vec{P}(t) = R(t) \vec{Q}(t) \qquad \text{with} \qquad
   R[\Theta(t)] = \begin{pmatrix}
                   \cos\Theta(t) & 0 & \sin\Theta(t) \\ 0 & 1 & 0 \\ -\sin\Theta(t) & 0 & \cos\Theta(t)
                  \end{pmatrix},
\ee
where $\vec{Q}$ is the coherence vector in the new {\it co-rotating frame} and $R(t)$ is the time-dependent
rotation matrix. The QKE in the new basis appear as
\be
   \frac{\partial}{\partial t} \vec{Q}(t) = S_Q(t) \vec{Q}(t), 
   \qquad S_Q = \begin{pmatrix}
                 0 & -\lambda \cos\Theta - \beta\sin\Theta & -\frac{\mathrm d \Theta}{\mathrm d t} \\
                 \lambda \cos\Theta + \beta\sin\Theta & 0 & -\beta\cos\Theta + \lambda\sin\Theta \\
                 \frac{\mathrm d \Theta}{\mathrm d t} & \beta\cos\Theta - \lambda\sin\Theta & 0
                \end{pmatrix}.
\ee 
Since we introduced $\Theta$ as a generic time-dependent mixing angle we are endued with its explicit definition
according to our needs. It is readily seen that the $\left(S_Q\right)_{23}$ and $\left(S_Q\right)_{32}$ elements of the evolution matrix can be eliminated by an appropriate choice of the mixing angle $\Theta(t)$. The advantage of this choice is the geometrical interpretation: in the $Q$-picture the motion of the coherence vector is confined to the $xy$-plane, if there was not the additional perturbation by the time derivative of the effective angle, which introduces a non-zero $z$-component to the problem and forces the motion to exit the $xy$-plane as the ensemble evolves. The smaller the change of the effective mixing with time, the smaller the urge of the coherence vector to exit the $xy$-plane. Therefore we fix the effective mixing angle to be
\be
   \cos\Theta(t) = \frac{\lambda(t)}{\sqrt{\lambda^2(t) + \beta^2(t)}}, \qquad 
   \sin\Theta(t) = \frac{\beta(t)}{\sqrt{\lambda^2(t) + \beta^2(t)}}.
\ee
The effective mixing angle reveals that mixing becomes maximal $(\Theta = \pi/2)$ if the condition
\be
   \lambda(t_{\text{res}}) = 0
\ee
is satisfied for the so introduced resonant time $t_{\text{res}}$. A vanishing $\lambda(t)$, i.e. maximal effective mixing, hence coincides with the existence of a resonance in neutrino conversions, which can also be equivalently
rephrased for a resonant temperature $T_{\text{res}}$, depending on the application one has in mind.
\par
We now recast the evolution matrix in the $Q$-picture as
\be
   S_Q = \begin{pmatrix}    
          0 & -\omega_{\text{eff}} & -\frac{\mathrm d \Theta}{\mathrm d t} \\
          \omega_{\text{eff}} & 0 & 0 \\
          \frac{\mathrm d \Theta}{\mathrm d t} & 0 & 0
         \end{pmatrix}.
\ee
Consider the matrix norm of this evolution matrix in the new quantum mechanical picture
\be
   \vert\vert S_Q \vert\vert_{\mathrm F}^2 = 2 \omega^2_{\text{eff}} \left[1 + 
   \left(\frac{1}{\omega_{\text{eff}}}\frac{\mathrm d \Theta}{\mathrm d t}\right)^2\right]. 
\ee
At first glance the above transformation seems to worsen the convergence properties due to the appearance of the
additional
\be
   \gamma \equiv \frac{1}{\omega_{\text{eff}}}\frac{\mathrm d \Theta}{\mathrm d t} \label{gamma}
\ee
term. However, if this very term is sufficiently small, $\gamma \ll 1$, the convergence will only be marginally altered. Moreover, the smallness condition can be understood physically as well: the characteristic time scale
of the system under study is $\tau_{\text{sys}} = 1/ \omega_{\text{eff}}$, whereas the characteristic time scale of
the interaction can be identified as $\tau_{\text{int}} = (\mathrm d \Theta/ \mathrm d t)^{-1}$. Hence the parameter
$\gamma$ simply compares the characteristic time scale of the system to the characteristic time scale of the interaction, stating that a small $\gamma$ can be paraphrased as the system's time scale being much smaller than the interaction's time scale. Put another way, the interaction is adiabatic. The parameter $\gamma$ is thus readily interpreted and henceforth referred to as the adiabaticity parameter for the system.
\par
The evolution matrix of the system thus reads
\be
   S_Q = \begin{pmatrix}
          0 & -\omega_{\text{eff}} & -\gamma\omega_{\text{eff}} \\
          \omega_{\text{eff}} & 0 & 0 \\
          \gamma\omega_{\text{eff}} & 0 & 0
         \end{pmatrix}.
\ee
Before we move on with our analysis it is just to briefly comment on the definition of the 
adiabaticity parameter and the effective mixing angle. The effective mixing angle defined above is essentially the
expression encountered when it comes to the usual MSW framework of matter-affected neutrino oscillations. Note, however, that the latter typically features a $\sin2\Theta$ instead of $\sin\Theta$ as defined here. In order to streamline notation we will nonetheless still omit this factor of two. Moreover, we alert the reader that defining
$1/\gamma$ as the adibaticity parameter is also quite common in the literature. However, the physics is not altered by this convention. Also when comparing our analysis to other work it is important to notice that in the $\Theta$ instead of $2\Theta$ convention the adiabaticity parameter lacks a factor of two as well. We will analyze the adiabaticity parameter further in section \ref{nad}.
\par
It is by now established that $\gamma$ can serve as a small perturbation parameter in the adiabatic regime of neutrino
conversions. It thus feels harmonious to struggle through just another transformation which is introduced to isolate
the perturbation parameter $\gamma$ in a convenient way and such that fast convergence of the expansion to come is 
assured. The prescription is as follows
\be
   \vec{Q}(t) = U(t) \vec{X}(t), \qquad \text{where} \qquad \frac{\partial}{\partial t}U(t) = S_Q^{\omega}(t) U(t),
                                 \qquad U(t_0) = {\bf 1},
\ee
having decomposed the evolution matrix according to
\be
   S_Q = S_Q^{\omega} + S_Q^{\gamma} = 
                                       \begin{pmatrix}    
                                        0 & -\omega_{\text{eff}} & 0 \\
                                        \omega_{\text{eff}} & 0 & 0 \\
                                        0 & 0 & 0
                                       \end{pmatrix}
                                      +\begin{pmatrix}    
                                        0 & 0 & -\gamma\omega_{\text{eff}} \\
                                        0 & 0 & 0 \\
                                        \gamma\omega_{\text{eff}} & 0 & 0
                                       \end{pmatrix}
\ee
in self-obvious notation. The subsidiary evolution equation for $U(t)$ is also readily solved to give
\be
   U(t) = \begin{pmatrix} 
           \cos\tilde\omega_{\text{eff}} & -\sin\tilde\omega_{\text{eff}} & 0 \\
           \sin\tilde\omega_{\text{eff}} & \cos\tilde\omega_{\text{eff}} & 0 \\
           0 & 0 & 1
          \end{pmatrix}
   \qquad \text{and} \qquad \tilde\omega_{\text{eff}}(t) = \int_{t_0}^t \mathrm d \tau ~\omega_{\text{eff}}(\tau).
\ee
The QKE in disguise are recognized to be
\be
   \frac{\partial}{\partial t}\vec{X}(t) = S_X(t)\vec{X}(t), 
   \qquad S_X(t) = \begin{pmatrix} 
                    0 & 0 & -\gamma\omega_{\text{eff}}\cos\tilde\omega_{\text{eff}} \\
                    0 & 0 & \gamma\omega_{\text{eff}}\sin\tilde\omega_{\text{eff}} \\
                    \gamma\omega_{\text{eff}}\cos\tilde\omega_{\text{eff}} & -\gamma\omega_{\text{eff}}\sin\tilde\omega_{\text{eff}} & 0
                   \end{pmatrix}
\ee
and calculating the matrix norm  yields 
\be
   \vert\vert S_X \vert\vert_{\mathrm F}^2 = 2 \gamma^2\omega_{\text{eff}}^2.
\ee
It is evident now that the small parameter in the adiabatic regime, namely $\gamma$, has been isolated
and hence good convergence of the sought-after perturbation theory can be expected. This comforts us to seek 
a perturbative expansion in this basis (which is but an interaction picture for the $\vec{Q}$-basis).
\par
The considerations unfolded in this section have seen two linear transformations $R(t)$, $U(t)$ from the original
$\vec{P}$-basis to the $\vec{Q}$- and $\vec{X}$-basis. The reason for those transformations is twofold:
On the one hand changing the basis for the QKE discloses the physics of the system we are dealing with and on the other hand it seems advisable to find a basis for the QKE in which an approximate solution gives accurate results.
For convenience we shall now recapitulate the meaning of the transformations introduced so far.
\par
The first transformation ($\vec{P} \stackrel{R}{\to} \vec{Q}$) is inherently physical. It gives a recipe how to 
establish the concept of a {\it mass eigenbasis} in the coherence vector description of neutrino oscillations. The effective mixing angle defined in this way differs from the effective mixing angle encountered in the common MSW formalism by a conventional factor of two. In this quantum mechanical
picture a clear path of approaching the resonance in neutrino oscillations is unveiled. A resonant conversion of neutrino flavors is encountered for $\lambda(t) = 0$. Moreover, the transformation to the matter eigenbasis sees
the introduction of an effective mixing angle, which is a harbinger for the adiabaticity parameter $\gamma$ subsequently layed open. Adiabatic neutrino conversion occurs for $\gamma \ll 1$, when the time scale of the system
is much smaller than the time scale of the interaction. The mathematical benefit of this transformation is that we get a grasp on the convergence properties of the approximation we want to employ and we can furnish it with physical meaning. The convergence properties of the expansion get worse as the amount of adiabaticity violation increases. A fact which later will be useful to construct a non-adiabatic perturbation theory. Also the change to the matter eigenbasis suggests that the adiabaticity parameter $\gamma$ should be the appropriate small quantity to expand in.
\par
The second transformation ($\vec{Q} \stackrel{U}{\to} \vec{X}$) is convenient from a mathematical point of view. It
removes an exactly integrable part of the evolution matrix and thus the matrix norm for the latter is directly
proportional to the small expansion parameter $\gamma$. This truly renders $\gamma$ into the sought-after perturbation
parameter and we take comfort that the envisaged expansion converges fast. Put another way, already the first approximant should provide good guidance for the exact solution.
\par
Note eventually that no attempt for solving the QKE has been made so far. We have merely changed the quantum mechanical pictures to unfold the underlying physics. The paradigm of our analysis is that a careful treatment of the QKE, i.e. a succession of different changes of the quantum mechanical pictures already allows to extract important information about the system under consideration without ever explicitly solving the QKE.
\par\noindent
\newline
In the $\vec{X}$-basis the QKE are solved to first order in the Magnus expansion by
\be
   \vec{X}^{(1)}(t) = \text{exp}\left[\int_{t_0}^t \mathrm d \tau ~S_X(\tau)\right]\vec{X}(t_0).
\ee
The formal solution for the coherence vector $\vec{P}(t)$ to first order in the Magnus expansion is thus obtained 
as
\be
   \vec{P}^{(1)}(t) = R(t) U(t) ~\text{exp}\left[\int_{t_0}^t \mathrm d \tau ~S_X(\tau)\right]~R^{-1}(t_0)\vec{P}(t_0).
\ee 
In order to streamline notation we write the terms contained in the matrix exponential as
\be
   J_{s}(t) &=& \int_{t_0}^t \mathrm d \tau ~ \gamma\omega_{\text{eff}} \sin\tilde\omega_{\text{eff}}, \\
   J_{c}(t) &=& \int_{t_0}^t \mathrm d \tau ~ \gamma\omega_{\text{eff}} \cos\tilde\omega_{\text{eff}}
\ee 
as well as
\be
   \left|J\right| \equiv \sqrt{J^2_c + J^2_s}
\ee 
and the resultant expression for the coherence vector assumes a form
\be
   \vec{P}^{(1)}(t) = 
 &&\begin{pmatrix}
    \cos\Theta(t) & 0 & \sin\Theta(t) \\ 0 & 1 & 0 \\ -\sin\Theta(t) & 0 & \cos\Theta
   \end{pmatrix}
   \begin{pmatrix} 
    \cos\tilde\omega_{\text{eff}} & -\sin\tilde\omega_{\text{eff}} & 0 \\
    \sin\tilde\omega_{\text{eff}} & \cos\tilde\omega_{\text{eff}} & 0 \\
    0 & 0 & 1
   \end{pmatrix} \times \nonumber \\
 &&\times \frac{1}{\left|J\right|^2} 
   \begin{pmatrix}
   J^2_s + J^2_c \cos\left|J\right| & -J_c J_s \left(-1+\cos\left|J\right|\right)
   & -J_c \left|J\right|^2 \text{sinc} \left|J\right| \\
   -J_c J_s \left(-1+\cos\left|J\right|\right) & J^2_c + J^2_s \cos\left|J\right| & 
   J_s \left|J\right|^2 \text{sinc} \left|J\right| \\
   J_c \left|J\right|^2 \text{sinc} \left|J\right| & -J_s \left|J\right|^2 \text{sinc} \left|J\right| & \left|J\right|^2 \cos\left|J\right|
   \end{pmatrix} \times \nonumber \\
   &&
   \times
   \begin{pmatrix}
   \cos\Theta_0 & 0 & -\sin\Theta_0 \\
   0 & 1 & 0 \\
   \sin\Theta_0 & 0 & \cos\Theta_0
   \end{pmatrix} \vec{P}^i, \label{Pad}
\ee
where additionally $\Theta(t_0) \equiv \Theta_0$ and $\text{sinc} x \equiv \frac{\sin x}{x}$ was defined. This expression, as cumbersome as it may look at a first glance, presents an analytic, yet perturbative, solution to the QKE as given in Eq.(\ref{QKEII}) for a generic
potential, i.e. for a generic time (or equivalently temperature in early universe applications) dependence of both
$\beta$ and $\lambda$, as long as the transition can be considered adiabatic ($\gamma \ll 1$). Note also that oscillation probabilities in the one particle interpretation can be extracted from this formal solution. Moreover, oscillating contributions to this very probability can be studied since there is no inherent averaging over rapidly oscillating contributions as is usually considered in the derivation of the oscillation probability in the MSW framework. Still, to fully appreciate this result a thorough discussion of various limiting cases, such as the adiabatic limit, is called for \cite{Ioannisian:2008ve}.
We postpone this endeavor until section \ref{limit}.

\section{Non-adiabatic perturbation theory}\label{nad}

Before we proceed by developing a non-adiabatic perturbation theory on the same grounds as the foregoing adiabatic
perturbation theory it is instructive to briefly reconsider the adiabaticity parameter as defined in Eq.(\ref{gamma})
and rephrase it in a way that allows for an understanding of the notion of adiabaticity in terms of the
parameters $\beta$ and $\lambda$. It is straightforward to show that
\be
   \gamma(t) = \frac{\beta\lambda}{\omega^3_{\text{eff}}} ~\frac{\mathrm d}{\mathrm d t} \ln \frac{\beta}{\lambda}.
\ee
However, physically the adiabaticity parameter at the neutrino conversion resonance ($\lambda = 0$) is of foremost interest. We find
\be
   \gamma_{\text{res}} \equiv \gamma(t_{\text{res}}) = - \frac{1}{\beta^2(t_{\text{res}})} 
                       \left.\frac{\mathrm d \lambda}{\mathrm d t}\right|_{t = t_{\text{res}}}.
\ee
The adiabaticity parameter depends on the shape of the matter profile $\mathrm d \lambda / \mathrm d t$ as is
expected from the MSW framework. Large variations of the matter profile at resonance are clearly disfavored for the sought-after perturbation expansion to work. Besides this contribution also the term $1/\beta^2$ is familiar. It simply states that the larger $\beta$, the better the expansion works. Put another way, if the notion of adiabaticity
as put forward in our analysis is adopted, a small $\beta$ at resonance is incompatible with an adiabatic perturbation
expansion to some extent. A regime with small $\beta$ thus indicates non-adiabatic transitions; we will also refer to this regime as the sudden regime henceforth.
\par\noindent
\newline
Recalling the QKE according to Eq.(\ref{QKE}) it is obvious that $\beta$ itself can be adopted as a small perturbation parameter if a non-adiabatic perturbation theory is desired. We split the evolution matrix $S$
as
\be
   S(t) = S_{\lambda}(t) + S_{\beta}(t) = 
        \begin{pmatrix}
         0 & -\lambda(t) & 0 \\ 
         \lambda(t) & 0 & 0 \\
         0 & 0 & 0
        \end{pmatrix}
        +
        \begin{pmatrix}
         0 & 0 & 0 \\ 
         0 & 0 & -\beta(t) \\
         0 & \beta(t) & 0
        \end{pmatrix}
\ee
in obvious notation. The $S_{\lambda}$ subsystem of this evolution equation can be integrated exactly and hence
we impose the following change of the quantum mechanical picture
\be
   \vec{P}(t) = V(t)\vec{Y}(t) \qquad \text{with} \qquad \frac{\partial}{\partial t}V(t) = S_{\lambda}(t) V(t).
\ee
The subsidiary evolution equation again is solved to give
\be
   V(t) = \begin{pmatrix} 
           \cos\tilde\lambda & -\sin\tilde\lambda & 0 \\
           \sin\tilde\lambda & \cos\tilde\lambda & 0 \\
           0 & 0 & 1
          \end{pmatrix}
   \qquad \text{and} \qquad \tilde\lambda(t) = \int_{t_0}^t \mathrm d \tau ~\lambda(\tau),
\ee
rephrasing the QKE as
\be
   \frac{\partial}{\partial t} \vec{Y}(t) = S_Y(t) \vec{Y}(t), \qquad 
   S_Y(t) = \begin{pmatrix}
             0 & 0 & -\beta\sin\tilde\lambda \\
             0 & 0 & -\beta\cos\tilde\lambda \\
             \beta\sin\tilde\lambda & \beta\cos\tilde\lambda & 0
            \end{pmatrix}.
\ee
Note also, that the new quantum mechanical picture is just the interaction picture. The matrix norm reveals isolation
of the small perturbation parameter:
\be
   \vert\vert S_Y \vert\vert^2_{\mathrm F} = 2\beta^2.
\ee 
\par
Solving the QKE for $\vec{Y}(t)$ will see the time-integrated evolution matrix $S_Y$ since the first order Magnus 
expansion gives the coherence vector as
\be
   \vec{P}^{(1)}(t) = V(t) ~\text{exp}\left[\int_{t_0}^t \mathrm d \tau ~S_Y(\tau)\right] ~\vec{P}(t_0)
\ee
and it is thus sensible to define the following integrals to streamline notation
\be
   K_s(t) &=& \int_{t_0}^{t} \mathrm d \tau ~\beta\sin\tilde\lambda, \\
   K_c(t) &=& \int_{t_0}^{t} \mathrm d \tau ~\beta\cos\tilde\lambda
\ee
as well as 
\be
   \left|K\right| \equiv \sqrt{K_{c}^2 + K_{s}^2}
\ee
in order to mimic the notation introduced above for the adiabatic perturbation theory.
The coherence vector to first order in the Magnus expansion is calculated to be
\be
   \vec{P}^{(1)}(t) =
   &&\begin{pmatrix}
   \cos\tilde\lambda & -\sin\tilde\lambda & 0 \\
   \sin\tilde\lambda & \cos\tilde\lambda & 0 \\
   0 & 0 & 1
   \end{pmatrix} \times \label{cohvecapproxnonad}  \nonumber \\
   &&
   \times \frac{1}{\left|K\right|^2}
   \begin{pmatrix}
   K_{c}^2 + K_{s}^2 \cos\left|K\right| & K_{c}K_{s} \left(-1 + \cos\left|K\right|\right)
   & -K_{s}\left|K\right|^2 \text{sinc}\left|K\right| \\
   K_{c}K_{s} \left(-1 + \cos\left|K\right|\right) & K_{s}^2 + K_{c}^2 \cos\left|K\right| & 
   -K_{c} \left|K\right|^2 \text{sinc}\left|K\right| \\
   K_{s} \left|K\right|^2 \text{sinc}\left|K\right| & K_{c}\left|K\right|^2 \text{sinc}\left|K\right| & 
   \left|K\right|^2 \cos\left|K\right|
   \end{pmatrix} \vec{P}^i .\label{Pnad}
\ee
This is the analytic perturbative solution to the QKE in Eq.(\ref{QKEII}) with generic potential and time
dependence for $\beta$ and $\lambda$ as long as the evolution can be considered non-adiabatic, which is 
equivalent to saying that $\beta$ is a small quantity one can expand in. Again, this result can only be fully
appreciated once the associated limiting cases are recovered. We will see to this in the next section 
\cite{D'Olivo:1990xs}.

\section{Perturbation theory ingredients and limiting cases}\label{limit}

We understand that our approach is a generic solution to the QKE in Eq.(\ref{QKEII}) without making any explicit reference to the physics it can be applied to. Hence the integrals $J_{c/s}(t)$, $K_{c/s}(t)$ explicitly depend
on the time dependence of both $\lambda$, i.e. the potential term $V_{\alpha}$, as well as $\beta$ and have to be evaluated in each application separately. However, certain general statements can be made already due to the fact
that neutrino conversions reveal a resonance at $\lambda = 0$. 
\par
In any case, however, it is still necessary to demonstrate that 
the Magnus expansion does give exact results in the various physical limits.

\subsection{The integrals: $J_{c/s}(t)$ and $K_{c/s}(t)$}\label{limitint}

The structure of the integrals $J_{c/s}(t)$ and $K_{c/s}(t)$, for the adiabatic and non-adiabatic case, 
respectively, at a first glance, is similar: the integrand is the 
expansion parameter ($\gamma\omega_{\text{eff}}$ for the adiabatic case; $\beta$ for the non-adiabatic case) multiplied by an oscillatory function. On second thought, however, there is a crucial difference; the oscillatory
term in $K_{c/s}$ has a stationary phase ($\mathrm d \tilde\lambda / \mathrm d t = 0$ at resonance), whereas 
$J_{c/s}$ does not.
\par
Let us therefore evaluate $K_{c/s}$ by means of the stationary phase method. Two main assumptions are needed to apply
the stationary phase method: the oscillatory behavior of the integrand must be rapid enough to suppress all large
contributions to the integral which might come from $\beta$ so that the latter can simply be evaluated at resonance. The other requirement is that the resonance happens in a {\it small} region around $t_{\text{res}}$; put another way,
the smallness of the aforementioned region is determined by whether the substitution $t_0 \to -\infty$ and
$t \to \infty$ is justified in this region or not. Supposing that these two requirements are met, we find 
\be
   \left|K_{s}\right| &=& 0, \\
   \left|K_{c}\right| &\simeq& \left[\frac{2\pi}{\gamma_{\text{res}}}\right]^{1/2}, \\
   \text{hence} \qquad\qquad \left|K\right| &\simeq& \left[\frac{2\pi}{\gamma_{\text{res}}}\right]^{1/2}.
\ee
Note, that in the non-adiabatic perturbation theory the reciprocal value of the adiabaticity parameter at resonance
is a small quantity.
\par
Consider the integrals $J_{c/s}$ now. The first step that comes to mind here is integration by parts. We get
\be
   J_{s}(t) &=& -\gamma\cos\tilde\omega_{\text{eff}}\Big|_{t_0}^t + \int_{t_0}^t \mathrm d \tau ~
              \frac{\mathrm d \gamma}{\mathrm d \tau} \cos\tilde\omega_{\text{eff}}, \\
   J_{c}(t) &=& ~~~\gamma\sin\tilde\omega_{\text{eff}}\Big|_{t_0}^t - \int_{t_0}^t \mathrm d \tau ~
              \frac{\mathrm d \gamma}{\mathrm d \tau} \sin\tilde\omega_{\text{eff}}.
\ee
If the variation of $\gamma$ in the interval $[t_0,t]$ is sufficiently mild the main contribution to the 
integrals is expected to come from the first term on the right hand side.
\par
Given these arguments the integrals $K_{c/s}$, $J_{c/s}$ reveal a common trademark. Both integrals turn out to
be small in the following sense: $K_{c/s}$ is proportional to the inverse of $\gamma_{\text{res}}$ which, in a
non-adiabatic perturbation theory is a large quantity; likewise in the adiabatic perturbation theory $\gamma$
is the small quantity to expand in and again the integrals $J_{c/s}$ turn out to be proportional to $\gamma$.

\subsection{Limiting cases}\label{limitcases}

We begin our consideration by studying appropriate limits for the adiabatic perturbation theory first.
 \begin{enumerate}
    \item {\it Adiabatic perturbation theory: The vacuum limit} \\ 
          This limit to Eq.(\ref{Pad}) is probably the most intuitive one, if we confine ourselves to the 
          one particle interpretation. We understand that for an exactly solvable system the first order Magnus term
          should already give the exact result, hence implying $\vec{P}^{(1)}(t) \equiv \vec{P}(t)$. 
          Let us examine how this works out here: Firstly, we discard
          the potential term $V_{\alpha}$. This gives $\lambda \to -\frac{\Delta m^2}{2p} \cos2\theta_0$ and
          $\beta = \frac{\Delta m^2}{2p} \sin2\theta_0$. 
          Hence, for consistency we must take the adiabatic limit of $\gamma \to 0$, which immediately implies
          $J_{c/s} \to 0$. Moreover, $\tilde\omega_{\text{eff}} \to \omega t$ if we define the common oscillation
          frequency $\omega = \frac{\Delta m^2}{2p}$ and set $t_0 = 0$ (as no resonance time exists, the choice of
          $t_0$ is arbitrary). Finally, it follows directly from the definition of the effective mixing angles
          that $\Theta(t) \to 2\theta_0$ for all times.
          All this reduces the coherence vector to
          \be
             \vec{P}^{(1)}(t) =
             \begin{pmatrix}
             \cos^22\theta_0 \cos\omega t + \sin^22\theta_0 & \cos2\theta_0\sin\omega t
             & \sin4\theta_0 \sin^2\frac{\omega t}{2} \\
             \cos2\theta_0\sin\omega t & \cos\omega t & -\sin2\theta_0\sin\omega t \\
             \sin4\theta_0 \sin^2\frac{\omega t}{2} & -\sin2\theta_0\sin\omega t 
             & \cos^22\theta_0 + \sin^22\theta_0 \cos\omega t
             \end{pmatrix} \vec{P}^i \label{vaclimit}.
          \ee
          Suppose we start with a $\nu_a$ flavor such that $\vec{P}^i = (0,0,1)$. The probability to find the
          neutrino in the same/the other state after time $t$ is then using Eqs.(\ref{probaa}-\ref{probab}) 
          as well as Eq.(\ref{vaclimit}) given by
          \be
             \text{prob}(\nu_a \to \nu_a) &=& 1 - \sin^22\theta_0 \sin^2\frac{\omega t}{2}, \\
             \text{prob}(\nu_a \to \nu_b) &=&     \sin^22\theta_0 \sin^2\frac{\omega t}{2}
          \ee
          which is just the common oscillation probability. Note, moreover that this result was obtained
          solely using the truncated Magnus expansion as given above and that it accounts for probability 
          conservation. Put another way, unitarity is guaranteed by means of the expansion itself and does not have 
          to be imposed by hand.
 \end{enumerate}
We next turn our attention to the non-adiabatic case. There exist (at least) two interesting limits.
 \begin{enumerate}
    \item {\it Non-adiabatic perturbation theory: The sudden limit} \\
          The sudden limit of taking $\beta \to 0$ renders the QKE (\ref{QKEII}) into formally exactly solvable differential equations such that the Magnus expansion should give an exact result. The limit $\beta \to 0$ enforces $\omega_{\text{eff}} = \left|\lambda\right|$ and hence $\cos\Theta = 1$, $\sin\Theta = 0$, which in turn implies $\gamma \to 0$. The coherence vector assumes a form
          \be
             \vec{P}(t) = \begin{pmatrix}
                                   \cos\tilde\lambda & -\sin\tilde\lambda & 0 \\
                                   \sin\tilde\lambda & \cos\tilde\lambda & 0 \\
                                   0 & 0 & 1
                                \end{pmatrix} \vec{P}^{i}.
          \ee
          The coherences of the ensemble are oscillating as a function of time (the ensemble is incoherent) and the
          flavor is frozen to its initial value. Put another way, in physical situations in which the evolution of the ensemble happens in a way that with increasing time also $\beta$ increases, the unfreezing of the ensemble can be studied using non-adiabatic perturbation theory since it treats $\beta$ as a small perturbation. We will point out in section \ref{limitbeyond} that this is typically the case in early universe applications. There is, however, a twist when it comes to early universe applications in that such systems are typically collision-dominated at high temperatures and thus the notion of adiabaticity is expected to be modified due to the presence of collisions. Put another way, a small $\beta$ in early universe environments augmented by the presence of decohering collisions might as well allow for an adiabatic perturbation theory (see section \ref{limitbeyond} for some more details).
    \item {\it Non-adiabatic perturbation theory: The slab model limit} \\
          Suppose that we start the evolution of the neutrino ensemble from a purely $\nu_a$ state 
          $\vec{P}^i = (0,0,1)$. We obtain for the coherence vector
          \be
             \vec{P}^{(1)}(t) = \begin{pmatrix} (-K_{s} \cos\tilde\lambda + K_{c} \sin\tilde\lambda) 
                                ~\text{sinc}\left|K\right| \\
                                -(K_{c} \cos\tilde\lambda + K_{s} \sin\tilde\lambda) 
                                ~\text{sinc}\left|K\right| \\
                                \cos\left|K\right| 
             \end{pmatrix}.
          \ee
          The flavor oscillation probability is then written as
          \be
             \text{prob}(\nu_a \to \nu_a) = \cos^2\frac{1}{2}\left|K\right|.
          \ee
          Suppose furthermore that the situation as described above to estimate the $K$-type integrals holds,
          i.e. the resonance in neutrino conversions happens in a narrow time interval centered around 
          $t_{\text{res}}$. Applying the stationary phase approximation we then obtain
          \be
             \text{prob}(\nu_a \to \nu_a) = \cos^2\sqrt{\frac{\pi}{2}\frac{1}{\gamma_{\text{res}}}}
             %\left[\frac{\pi}{2}\frac{1}{\gamma_{\text{res}}}\right]^{1/2}.
          \ee
          This result is the oscillation probability for the slab model as outlined in 
          Ref.~\cite{Rosen:1986jy}. So the relevant limit is respected in this situation.
\end{enumerate}

\subsection{Higher order corrections and applications}\label{limitbeyond}

As has been seen in the previous sections the Magnus expansion can be easily extended to higher orders by summing
the associated approximants according to $\Omega(t) = \Omega_{1}(t) + \Omega_{2}(t) + \dots$. In order to get a grasp on how this prescription unfolds we calculate the 2nd order Magnus approximant to be
\be
   \Omega_2(t) = \begin{pmatrix} 0 & \mathcal{J} & 0 \\ -\mathcal{J} & 0 & 0 \\ 0 & 0 & 0 \end{pmatrix},
\ee
where $\mathcal{J}$ is given by
\be
   \mathcal{J}_{\text{ad}} = \frac{1}{2}\int_{t_0}^{t}\mathrm d t_1 \int_{t_0}^{t_1} \mathrm d t_2 
                 ~\gamma(t_1)\gamma(t_2)\omega_{\text{eff}}(t_1)\omega_{\text{eff}}(t_2) \sin\left[\tilde\omega_{\text{eff}}(t_2)-\tilde\omega_{\text{eff}}(t_1)\right]
\ee
for the adiabatic case and
\be
   \mathcal{J}_{\text{nad}} = \frac{1}{2}\int_{t_0}^{t}\mathrm d t_1 \int_{t_0}^{t_1} \mathrm d t_2 
                 ~\beta(t_1)\beta(t_2)\sin\left[\tilde\lambda(t_2)-\tilde\lambda(t_1)\right]
\ee
for the non-adiabatic case respectively. The calculations are performed in the $X$-picture for adiabatic transitions
and in the $Y$-picture for non-adiabatic transitions. Two things are readily inferred: the second order approximant is indeed $\mathcal{O}(\gamma^2)$ and $\mathcal{O}(\beta^2)$ for adiabatic and non-adiabatic corrections respectively as is expected. Moreover it is seen that the 2nd order populates the $(23)$ and $(32)$ entries
of the Magnus operator $\Omega$. 
\par\noindent
\newline
We shall now shortly elaborate on the complications which arise when ensemble decoherence is to be taken into account.
This typically happens in early universe applications in which the time evolution of neutrinos is governed by three
distinct physical processes: Firstly, the expansion of the universe. Secondly, coherent oscillations governed by a matter-dependent effective Hamiltonian which results from coherent forward scattering processes of neutrinos off the background particles. Thirdly, scattering processes with the background plasma of elementary particles.
These collisions, or non-forward scattering processes, with particles from the background medium typically
introduce decoherence effects into the neutrino ensemble. In our analysis in this paper we have neglected the ensemble decoherence due to non-forward scattering.
\par
The epoch of foremost interest in studying neutrino oscillations in the early universe is the one between muon decoupling
at $T \sim m_{\mu} \sim 100 ~\text{MeV}$ and neutrino decoupling, i.e. prior to Big Bang Nucleosynthesis (BBN),
at about $T \sim 1 ~\text{MeV}$, since during this time the initial conditions for Nucleosynthesis, the electron
neutrino abundance, are set~\footnote{Note, however, that an additional restriction in the derivation of the
QKE emerges. If a treatment using the $S$ matrix is employed, it is understood that the initial and final
scattering states evolve as ``free states''. This is only the case if the plasma is sufficiently dilute and the
quanta do not spend most of their time interacting. For the case of a weakly interacting gas of relativistic
particles a rough estimate yields that the approach using the $S$ matrix should be valid for 
temperatures $T \ll 100 ~\text{GeV}$, which does not interfere with the temperature regime under consideration
here.} 
which then directly influence the neutron-to-proton ratio at the onset of BBN via
$\beta$-processes $p + e^- \rightleftharpoons n + \nu_e$. The primordial plasma during this epoch thus consists of 
electrons, positrons, neutrinos and antineutrinos.
\par
The density matrix $\rho$ for the system of interacting and oscillating neutrinos encodes ratios of number density
distributions and hence the expansion of the universe does not directly contribute to the time evolution of 
the density matrix. However, the momenta of the particles are redshifted and the equilibrium number distributions $N^{\text{EQ}}(p,0)$ thus depend on time through this redshifting.
\par
So far most attention has been paid to two neutrino systems and specifically to active-sterile flavor oscillations
~\cite{Bell:1998ds}.
The latter are interesting since active-sterile oscillations would populate the additional sterile species and thus
contribute significant additional energy density, which in turn would trigger an accelerated expansion of the universe
and hence lead to a higher weak freeze-out temperature. This again would alter the neutron-to-proton ratio at the 
onset of BBN.
The coherent part of matter-affected active-sterile oscillations splits in two contributions. One is just the leading order density-dependent contribution (the Mikheev-Smirnov-Wolfenstein \cite{Wolfenstein:1977ue} (MSW) part). This part is only
temperature-dependent indirectly via the cosmological redshifting of fermion number density. The second contribution
comes from leading order finite temperature gauge boson effects~\cite{Notzold:1987ik}, which cannot be neglected
at the temperatures considered here.
\par
The loss of coherence is due to neutrino collisions with the background medium. The decoherence (or damping)
function for this process in thermal equilibrium turns out to be proportional to the total collision rate for the
neutrino with momentum $p$ under consideration. 
\par
The epoch of interest can now be decomposed into three distinct domains:
At high temperatures finite temperature gauge boson effects dominate. 
Repopulation effects from the background plasma can be neglected
since at high temperatures thermal equilibrium for all relevant species is rapidly established. 
At intermediate temperatures lepton number production starts and the
forward scattering contribution comes into play as a small perturbation. Finally prior to the onset of BBN, at low
temperatures, collisional effects and finite temperature gauge boson contributions cease to be important and
coherent neutrino oscillations are the dominant process.
\par
In each of the aforementioned temperature domains the QKE should be solved to determine the evolution of the
neutrino ensemble. It is obvious that decoherence and repopulation effects modify the underlying QKE and complicate
their analysis by introducing new physical scales in the system. The early universe framework thus deserves a more
careful treatment which is beyond the scope of this work.

\section{Conclusions}

We have developed a new formalism to solve the Quantum Kinetic Equations governing the evolution of a two-flavor
neutrino ensemble with finite temperature subject to generic potentials. To this end we use the Magnus expansion 
which has virtues when it comes to perturbatively solve coupled non-autonomous differential equations such as the
QKE considered here: the expansion is unitary to each order of the approximation and its convergence properties are determined by the matrix norm of the underlying evolution matrix, which depends on the physical parameters
of the system.
%\par
%The paradigm of our analysis bifurcates: 
%\par
%Starting from the QKE derived from the density matrix of the system we realize that the influence of the potential
%terms can be dealt with via a generic rotation which introduces the notion of an effective mixing angle. This mixing %angle adopts the property of being maximal at the neutrino conversion resonance. Put another way, the resonance in
%neutrino conversions can be defined in this way. Another benefit of the generic rotation about the $xz$-plane
%is that it allows to recast the QKE in a way that is readily interpreted geometrically. The motion of the neutrino %ensemble in coherence vector space is confined to the $xy$-plane and perturbation to adiabaticity force the %coherence vector to exit that plane as it evolves in time. The matrix norm of the evolution matrix in this %co-rotating picture
%allows to quantify the deviation from the oscillatory behavior in the $xy$-plane by introducing an adiabaticity %parameter of the system which compares the time scales of the system to the time scale of the interaction. Thus it
%is possible to define the notion of an effective mixing angle and adibaticity in a model-independent way by simply
%changing the quantum mechanical picture of the QKE. Eventually a change to the interaction picture isolates the 
%adiabaticity parameter in a way that proves convenient to perform a perturbative expansion of the solution for the coherence vector in this basis. 
The perturbation ansatz proposed maps the problem of diagonalizing a Hamiltonian with generic potentials (and hence possible time dependences) onto calculating the exponential of a matrix and the associated integrals as its
emergent ingredients.
\par
Contrary to previous approaches the formalism presented here describes both adiabatic and non-adiabatic flavor conversions in a quantum neutrino
ensemble at finite temperatur and arbitrary (e.g. matter) potentials. In the adiabatic case a suitable succession of bases changes is employed to define an effective oscillation frequency, mixing angle and adiabaticity parameter for the ensemble.
The latter is then used as a perturbation parameter. In the non-adiabatic case the results obtained for the adiabaticity parameter are used to identify its inverse at the resonance as a suitable expansion parameter.
We give explicit expressions for the coherence vector to first order perturbation theory for both adiabatic 
and non-adiabatic conversions in the ensemble. We show that our approach is compatible with the appropriate limiting
cases for both adiabatic and non-adiabatic conversions and elaborate on how to tackle the integrals emergent from the
perturbation theory.
\par
We understand that our analysis presents a promising base for further investigations concerning the inclusion of
decoherence in the ensemble and the resulting applications in early universe scenarios, e.g. the generation of lepton number in active-sterile oscillations prior to BBN. Moreover, the approach may be useful for the analysis of long-standing oscillation anomalies such as MiniBooNE or even MSW-like scenarios in astrophysical environments.
%\par
%Exploiting the adiabaticity parameter as put forward in our analysis at the neutrino conversion resonance
%allows to identify the non-adiabatic regime and an appropriate perturbation parameter therein. In this sudden regime
%the QKE are most conveniently solved with the Magnus expansion in the interaction picture.
%\par
%Eventually we elaborate on the emergent integrals and \textcolor{red}{whatever will be done in section %\ref{limitbeyond}}

\acknowledgments
We are grateful to Danny van Dyk and Thomas J. Weiler for discussions on the topic during the course of this work.

\end{document}